\documentclass[reprint,unsortedaddress,showkeys]{revtex4-1}

\usepackage[utf8]{inputenc}
\usepackage[T1]{fontenc}
\usepackage[english]{babel}

\usepackage{amsmath,amssymb,amsfonts,amsthm,dsfont}
\usepackage{graphicx}
\usepackage{hyperref}
\usepackage{braket}
\usepackage{tikz}

\usetikzlibrary{backgrounds,fit,decorations.pathreplacing}

\newtheorem*{prop}{Proposition}
\newtheorem*{theo}{Theorem}

\newcommand{\hilb}[1]{\mathcal{H}^{#1}}
\newcommand{\trc}[0]{\text{tr}}

\begin{document}


\title{Quantum state space-dimension as a quantum resource}

\author{A. Plastino}
    \affiliation{Instituto de Física La Plata (IFLP-CONICET),
    and Departamento de Física, Facultad de Ciencias Exactas,
    Universidad Nacional de La Plata, 115 y 49, C.C. 67,
    1900 La Plata, Argentina}

\author{G. Bellomo}
    \affiliation{Instituto de Física La Plata (IFLP-CONICET),
        and Departamento de Física, Facultad de Ciencias Exactas,
        Universidad Nacional de La Plata, 115 y 49, C.C. 67,
        1900 La Plata, Argentina}

\author{A.R. Plastino}
    \affiliation{CeBio and Secretaría de Investigaciones,
    Universidad Nacional del Noroeste de la
    Prov. de Buenos Aires (UNNOBA-CONICET),
    R. Saenz Peña 456, Junín, Argentina}

\date{May 26, 2015}

\begin{abstract}
We argue that the dimensionality of the space of quantum systems'
states should be considered as a legitimate resource for quantum
information tasks. The assertion is supported by the fact that
quantum states with discord-like capacities can be obtained from
classically correlated states in spaces of dimension large enough.
We illustrate things with some simple examples that justify our
claim.
\end{abstract}



\maketitle

The Hilbert-space dimension has been related to physical
resources for different physical systems, playing a fundamental
role in quantum computation. Basically, the idea is that ``if you
want to avoid supplying an amount of some physical resource that
grows exponentially with problem's size, the computer must be made
up of parts whose number grows nearly linearly with the number of
qubits required in an equivalent quantum computer. This thus
becomes a fundamental requirement for a system to be a scalable
quantum computer''~\cite{BKCD02}. Moreover, some recent results
show that quantum dimensionality could be regarded as a physical
entity. For example, Brunner et al. defined what they call
`dimension witnesses': observable quantities to estimate the
minimum dimension of a given system state-space consistent with a
number of measured correlations~\cite{Brun08,Gall10,Hend12}. In
the same spirit, Wehner et al. found a lower bound that gives a
fundamental limit on the dimension of the state to implement
certain measurement strategies~\cite{Wehn08}. Here, we propose to
consider the dimension of the Hilbert space as a legitimate
resource for quantum information processing. Our main argument
lies in the observation, due to Li and Luo~\cite{LiLu08}, that
quantum separable states can be obtained from reductions of
classically correlated ones. Although some authors have suggested
the possibility of understanding the size of the Hilbert space as
a resource by itself~\cite{MaDa13}, the assertion that it is a
quantum-better-than-classical resource was never technically
analyzed, as far as we know.

Under the discord paradigm, a classically  correlated state (or
simply, a classical state) is one that is information-wise
accessible to local observers. Given a discord-like measure
$\delta$ and a classical state $\sigma^{AB}$ of a composite
system ${A+B}$, one knows that ${\delta(\sigma^{AB})=0}$. The
following theorem, due to Li and Luo, demonstrates a notable
relation between separable states and classical
states~\cite{LiLu08}.
    \begin{theo} \label{theo:luo}
    A state $\rho^{ab}$, of a composite system ${a+b}$, is
    separable over ${\hilb{ab}=\hilb{a}\otimes\hilb{b}}$ if and
    only if there is a classical state $\sigma^{AB}$ over
    ${\hilb{AB}=\hilb{A}\otimes\hilb{B}}$, with
    ${\hilb{A}=\hilb{a}\otimes\hilb{\bar{a}}}$ and
    ${\hilb{B}=\hilb{b}\otimes\hilb{\bar{b}}}$, such that
        \begin{equation} \label{eq:luo}
        \rho^{ab}=\trc_{\bar{a}\bar{b}}[\sigma^{AB}] \,.
        \end{equation}
    Here, the state $\sigma^{AB}$ of the composite ${A+B}$
    should be regarded as a classical extension of the separable
    $\rho^{ab}$.
    \end{theo}
    
The proof is given by Li and Luo in Ref.~\cite{LiLu08}. The
next result follows directly from the above theorem:
    \begin{prop} \label{prop:class_oper}
    Any quantum task carried out using un-entangled states can
    also be undertaken using classically correlated states.
    \end{prop}
    
Indeed, if a quantum task needs appealing to a given un-entangled
state $\rho^{ab}$, then there exists a classical extension
$\sigma^{AB}$ from whose reduction $\rho^{ab}$ can be obtained
(Fig.~\ref{fig:scheme}). The scheme is straightforwardly generalized to
tasks requiring several input quantum states.
    \begin{figure}[t]
    \centerline{
    \begin{tikzpicture}[thick]
    \draw[decorate,decoration={brace,mirror},thick] (0,.7) to
    node[midway,left] (bracket)
        {$\sigma^{AB}\,$(classical)$\;$} (0,-.7);
    \node at (.3,+.6) (a1) {$\bar{a}$};
    \node at (.3,+.2) (a2) {$a$};
    \node at (.3,-.2) (b1) {$b$};
    \node at (.3,-.6) (b2) {$\bar{b}$};
    \node (cc) at (1.5,+.6) {} edge [-] (a1);
    \node (c1) at (1.5,+.2) {} edge [-] (a2);
    \node (c2) at (1.5,-.2) {} edge [-] (b1);
    \node (cc) at (1.5,-.6) {} edge [-] (b2);
    \draw[decorate,decoration={brace},thick] (1.6,.3) to
    node[midway,right] (bracket)
        {$\;\rho^{ab}\,$(separable)} (1.6,-.3);
    \end{tikzpicture}
    }
    \caption{Every quantum state that is separable (within a
    given bipartition of the full system) is, in a formal sense,
    the reduction of a classical state of a system defined over
    a larger state-space (and preserves the original bipartition).}
    \label{fig:scheme}
    \end{figure}

Un-entangled quantum correlations, discord-ones  in particular,
have proved their usefulness both in the interpretation of
foundational quantum issues and in applications to quantum
information/computation  problems (see the excellent
review~\cite{Modi12,Stre14}).

We will explicitly illustrate here just how classically
correlated states can replace discord-possessing separable
states in two specific jobs: remote state preparation and
entanglement distribution.

\vspace{.6cm}\noindent
{\it Remote state preparation (RSP)}

\noindent
As a first illustration consider the RSP-protocol, a variant of
the well known teleportation-one, in which the  emitter knows the
state being sent to the recipient (for details see, for instance,
\cite{Daki12}). Daki\'c et al. showed that, for certain 2-qubits
states' family (those with  maximally-mixed marginals), the
protocol's fidelity coincides with the geometric discord of such
states. Girolami and Adesso singled out certain separable states
that maximize the geometric discord~\cite{GiAd11,GiAd11b}, although such
states do not possess maximally-mixed marginals. In fact, it is
easy to see that the RSP-fidelity for these states vanishes.
Instead, the state defined by the density matrix (standard basis)
    \begin{equation} \label{eq:rho_rsp}
        \rho_{RSP} = \frac{1}{4}
        \begin{pmatrix}
                    1 & 0 & 0 & 1 \\
                    0 & 1 & 0 & 0 \\
                    0 & 0 & 1 & 0 \\
                    1 & 0 & 0 & 1
        \end{pmatrix}
    \end{equation}
does maximize both the geometric  discord and the RSP-fidelity.
This state, defined in ${\hilb{a}\otimes\hilb{b}}$, can be
obtained (save for discord-preserving local unitary
transformations), as the reduction of the classical state (in
${\mathbb{C}^6\otimes\mathbb{C}^6}$)~\cite{BePP15}:
    \begin{equation} \label{eq:rho_ext}
        \sigma_{RSP} = \frac{1}{3}\sum_{i=1}^3{
            \ket{w_k,k}\bra{w_k,k}\otimes\ket{w_k,k}\bra{w_k,k} } \,,
    \end{equation}
with ${\ket{w_k,k}:=\ket{w_k}\otimes\ket{k}}$, respectively for $k=1,2,3$. ${\ket{w_k}:=\ket{\theta,\phi}}$,
${\ket{\theta,\phi}=\cos\left(\frac{\theta}%
{2}\right)\ket{0}+\exp(i\phi)\sin\left(\frac{\theta}{2}\right)}$,
where the pairs ${(\theta,\phi)}$ take the values ${(0,0)}$,
${(\frac{2\pi}{3},0)}$, and ${(\frac{2\pi}{3},\pi)}$. The states
$\ket{w_k}$ correspond to the  parties $a$ and $b$, while
${\{\ket{k}\}_{1\leq k\leq3}}$ are three orthogonal states in
${\mathbb{C}^3}$, corresponding to the extended parties $\bar{a}$
and $\bar{b}$. Thus, the same task can be performed with identical
efficiency by use of the classical extension.

Note that ${\rho_{RSP}}$ maximizes the geometric discord but not
the conventional one. This last discord, in
${\mathbb{C}^2\otimes\mathbb{C}^2}$, is maximized in the subset
of separable states by a different
states-family~\cite{GaGZ11,GiAd11,BePP15}.

\vspace{.6cm}\noindent
\textit{Entanglement distribution (ED)}

\noindent
Another example of the classical states' ability to perform
quantum tasks is that of \textit{entanglement distribution}~\cite{CVDC03}.
We use the following scheme. One starts with a
system in which two classically correlated, composite parties can
be identified, $A$ and $B$, represented by $\sigma^{AB}$. For $A$
we have the subparts $a$-$\bar{a}$, and for $B$, $b$-$\bar{b}$.
The reduction is ${\rho^{Ab}:=\trc_{\bar{b}}[\sigma^{AB}]}$. It
permits to tackle the job. In order to do so, consider two
partitions of the same state: ${ab|\bar{a}}$ is the initial
partition and ${a|\bar{a}b}$ the final one. Entanglement
distribution consists of the entanglement-increase in passing
from the initial to the final configurations. In such process,
the subsystem $b$ is taken
from being a partner of $a$ to being a partner of $\bar{a}$. So as to succeed,
the protocol does not need entangling  $A$ with $b$. Discord is
necessary in our partition, but not sufficient~\cite{CVDC03,StKB13}.

As an example, we start from a four-qubits classical state:
    \begin{equation} \label{eq:distrib_1}
    \gamma^{AB} = \sum_{k=1}^{4}{p_k\Pi^A_k\otimes\Pi^B_k} \,,
    \end{equation}
where ${{\{\Pi^A_k\}}_{1\leq k\leq4}}$ and
${{\{\Pi^B_k\}}_{1\leq k\leq4}}$ are basis of orthogonal, rank 1
projectors in $\mathbb{C}^4$, and ${{\{p_k\}}_{1\leq k\leq4}}$
is a probability distribution. So as to find ED-optimal classical
states, we generate random $\mathbb{C}^4$-basis for the $A$-$B$
parties. Specifically, we restrict our search to states such that
${p_k=\frac{1}{4}}$ and ${\Pi^A_k=\Pi^B_k}$ ${\forall k}$.
If ${E^{X|Y}}$ is the entanglement measure given by the negativity
in the partition ${X|Y}$ of the system, we find that $\gamma^{AB}$'s
ED is ${E^{a|\bar{a}b} - E^{ab|\bar{a}}\leq0.0915}$.

We now replace the initial state $\gamma^{AB}$ by another one in
which both $a$ and $b$ are composed by qu\textit{d}its, while
$\bar{a}$ and $\bar{b}$ retain their 2-qubits character. The new
state ${\gamma^{AB}_d}$ operates on
${\mathbb{C}^{2d}\otimes\mathbb{C}^{2d}}$, with ${d=1,2...}$. As
before, we can look for classical states maximizing  ED for each
value of $d$. We numerically did this for ${2\leq d\leq 6}$, again
restricting the search to states with $p_k=\frac{1}{2d}$ and
${\Pi^A_k=\Pi^B_k}$ ${\forall k}$. We encounter that the ED
augments with the dimension of the initial classical
state~(Tab.~\ref{tab:distrib}).

\begin{table}[!t]
    \setlength{\tabcolsep}{7pt}
    \centering
    \begin{tabular}{c c}
    \hline
    $d$         & Maximal ED \\
    \hline\hline
    2           & $ 0.0915 $ \\
    3           & $ 0.1269 $ \\
    4           & $ 0.1681 $ \\
    5           & $ 0.1744 $ \\
    6           & $ 0.3326 $ \\
    \hline
    \end{tabular}
    \caption{Maximal ED by classical states ${\gamma^{AB}_d}$ in
        ${\mathbb{C}^{2d}\otimes\mathbb{C}^{2d}}$.}
    \label{tab:distrib}
\end{table}

Accordingly, classically correlated states ${\gamma^{AB}_d}$
allow us to improve ED as long as we augment the Hilbert space
dimension.

\vspace{.6cm}\noindent
\textit{Work extraction from classical extensions}

\noindent
The distinction between classical and quantal can be made in
different ways. The discord establishes a division in the
capacity to locally `interrogate' a composite state. Another way,
introduced by Oppenheim et al.~\cite{Oppe02}, revolves around the work
that can be extracted from the state by quantum Maxwell demons~\cite{Zure03}.
If the whole work can be extracted by local demons, then the state
is classically correlated. The `work-deficit' between global and
local demons is a measure  of the correlations' `quantumness'.
Optimizing over all possible local measurements determines the
\textit{thermal discord}, which differs from the conventional
discord. The equivalence between information and
work~\cite{Szil29,Levi93} is of the essence to compare both types
of discord~\cite{Alic04,Horo05,Maru09,Hoso11}.

We have thus far shown that any separable state can be extended
to other, classically correlated states, and that such extensions
allow one to perform the same quantum tasks. From a
thermodynamical viewpoint, from the classical extension of a
given separable state one can always extract more work than from
the original quantum state. Indeed, we will now demonstrate as
an important new result, the validity of the relation
    \begin{equation} \label{eq:work}
    W^Q(\sigma^{AB}) = W^Q(\rho^{ab}) + W^Q(\rho^{aux}) + I(ab|aux) \,.
    \end{equation}
Here, ${\rho^{ab}=\sum_{k}p_k\rho^a_k\otimes\rho^b_k}$ ia a
quantum-correlated state, ${\sigma^{AB}}$ is a classical extension
of ${\rho^{ab}}$ (see Eq.~\eqref{eq:luo}) and
${\rho^{aux}:=\trc_{ab}{[\sigma^{AB}}]}$ the marginal state of
the ancilla. ${I(x|y):=S(x)+S(y)-S(x,y)}$ is the mutual quantum
information between the parties $x$ and $y$, with
${S(x):=-\trc[\rho^x\log_2\rho^x]}$.
${W^Q(\sigma^{AB}):= \log_2 d_{AB} - S(\sigma^{AB})}$ is the
maximum extractable work from ${\sigma^{AB}}$, when in contact
with a reservoir of temperature $T$, with $k_B$ Boltzmann's
constant, in units of ${k_BT=1}$. We determine in similar fashion
${W^Q(\rho^{ab})}$ and ${W^Q(\rho^{aux})}$.

Eq.~\eqref{eq:work} tells us that the extractable work from the
classical extension is bounded by below by the sum of the
extractable works from the original state plus that of and the
auxiliary part. This is simply demonstrated. It suffices to
appeal to the von Neumann entropy's sub-additivity and using it
in the preceding definition of ${W^Q(\sigma^{AB})}$.
Eq.~\eqref{eq:work} is important because it establishes a
relation between local (classical) resources and global (quantum)
ones, given that the whole extractable work from ${\sigma^{AB}}$
can be locally acceded (the thermal discord of ${\sigma^{AB}}$
vanishes in the partition ${A|B}$). This can be done, for instance,
by a local Maxwell demon (in $A$) that makes a measurement in
the eigen-basis of local projectors, ${\{\Pi^A_k\}}$, and
communicates y then with $B$, extracting work
${W^C(\sigma^{AB})=W^Q(\sigma^{AB})}$. Thus, from the above
equality it follows that
    \begin{equation} \label{eq:work2}
    W^C(\sigma^{AB}) \geq W^Q(\rho^{ab}) + W^Q(\rho^{aux}) \,.
    \end{equation}

Given that there exist a panoply of possible classical extensions
for a given separable $\rho^{ab}$, it makes sense to ask for the
optimal extension: that $\sigma^{AB}$ with the least possible
dimension~\cite{BePP15}. This kind of extension can be generally
encountered by recourse to the algorithm of Li and Luo. What is
peculiar in the extractable work from such optimal extension?

Consider the state ${\rho_{RSP}}$ of Eq.~\eqref{eq:rho_rsp}. In
addition to the classical extension given by ${\sigma_{RSP}}$
defined in ${\mathbb{C}^6\otimes\mathbb{C}^6}$
(Eq.~\eqref{eq:rho_ext}), one can also find an extension
${\tilde\sigma_{RSP}}$ in ${\mathbb{C}^8\otimes\mathbb{C}^8}$.
None of them is the optimal one. A Monte Carlo numerical search
suggests that the optimal extension, ${\sigma_{RSP}^{opt}}$,
acts on ${\mathbb{C}^4\otimes\mathbb{C}^4}$~\cite{BePP15}.
Computing the classical extractable works associated to each of
the extensions, one sees that the least-dimension extension is
the one that also minimizes the extractable work~(Tab.~\ref{tab:work}).
Such a result stimulates inquiry concerning whether the optimal
extension defined as the least dimension one does always coincide
with that of minimum extractable work.

The globally extractable work from a state ${\sigma^{AB}}$ is a
measure of the ability of distinguishing with respect to the
totally mixed state, since
${W^Q(\sigma^{AB}) = S(\sigma^{AB}||\mathds{1}^{AB}/d_{AB})}$,
with ${\mathds{1^{AB}}}$ the identity in $\hilb{AB}$. In this
sense, the classical extension of ${\rho^{ab}}$ that minimizes
the extractable work would be given by that state ${\sigma^{AB}}$
closest to ${\mathds{1^{AB}}/d_{AB}}$. The question remains
concerning whether the minimization of ${W^Q(\sigma^{AB})}$ (or
${S(\sigma^{AB}||\mathds{1}^{AB}/d_{AB})}$) is equivalent to the
minimization of ${d_{AB}}$, i.e., if both conditions indistinctly
determine the optimal classical extension.

\begin{table}[!t]
    \setlength{\tabcolsep}{7pt}
    \centering
    \begin{tabular}{c c c}
    \hline
    Extension               & Dimension & Extractable work \\
    \hline\hline
    $\sigma_{RSP}$          & 64        & $ 4.00 $ \\
    $\tilde\sigma_{RSP}$    & 36        & $ 3.58 $ \\
    $\sigma_{RSP}^{opt}$    & 16        & $ 2.00 $ \\
    \hline
    \end{tabular}
    \caption{Extractable work from different classical extensions
        of the separable state ${\rho_{RSP}}$. The
        minimum-dimension extension corresponds to minimum work.}
    \label{tab:work}
\end{table}

\vspace{.6cm}\noindent
\textit{Monogamy of correlations}

\noindent
Keeping in mind the equivalence between information and work
discussed above, relation~\eqref{eq:work2} can be regarded as a
monogamy one for the bi-partition ${ab|aux}$ of a classical
extension $AB$. The function
${i(\sigma^{AB}):=\log d_{AB} - S(\sigma^{AB})}$ can be seen as
the accessible information in the state $\sigma^{AB}$~\cite{Horo05}.
Thus, inequality \eqref{eq:work2} is equivalent to the inequality
${i(\sigma^{AB})\geq i(\rho^{ab}) + i(\rho^{aux})}$, that
determines a `hybrid' monogamous behavior concerning the classical
information of $\sigma^{AB}$ and the quantum information from
$\rho^{ab}$ and $\rho^{aux}$.

It is of great interest to find monogamy relations for different
measures of quantum and classical correlations. Except for
peculiar instances, quantum correlation measures do not satisfy
general monogamy relations. Even more, if an arbitrary measure
$Q$ possesses a few reasonable properties, it must vanish for
separable states in order to fulfill monogamy relations of the
type ${Q^{ab|c}\geq Q^{a|c}+Q^{b|c}}$~\cite{Stre12}. The usual
discord, for example, is not monogamous for general
states~\cite{Prab11,Gior11,Ren11,Stre12}.

The classical extensions that we are advancing here constitute a
clear example of monogamy violation because
${0=\delta^{ab|aux}(\sigma^{AB})\leq \delta^{a|aux}}$, and the
same holds for ${\delta^{b|aux}}$. All classical extensions
undergo discord-increase if some subsystem is discarded. Thus,
\textit{all classical extensions of separable states are
polygamous in the usual sense}. This observation i) constitutes
the basic argument in demonstrating that quantum correlations are
not monogamous in general~\cite{Stre12} and ii) underlies the
violation of more general monogamy relations, even for
multipartite correlation measures~\cite{Brag12,Liu14}. However,
there exist  monogamy relations valid even for classical
extensions if we consider some generalized multipartite quantum
correlations. Consider for instance the Global Quantum Discord
(GQD) --a symmetric discord extension for multipartite states--
defined as~\cite{RuSa11,Cele11} ${\delta_g(a_1|\dotsb|a_N):=%
\min_\Phi[I(a_1|\dotsb|a_N)-I^\Phi(a_1|\dotsb|a_N)]}$, where
${a_1,\dotsc,a_N}$ are the parties of an N-partite state
${\rho^{a_1\dotsb a_N}}$, where
${I(a_1|\dotsb|a_N):=\sum_k{S(a_k)}-S(a_1\dotsb a_N)}$ is the
generalized mutual information and  ${I^\Phi(a_1|\dotsb|a_N)}$ is
the mutual information after effecting a multilocal measurement
${\Pi^{\vec{\jmath}} := \{\Pi^{j_1}_{a_1} \otimes\dotsm\otimes%
\Pi^{j_N}_{a_N}\}}$, such that the post-measurement state becomes
${\Phi(\rho^{a_1\dotsb a_N}) = \sum_{\vec{\jmath}}%
{\Pi^{\vec{\jmath}} \rho^{a_1\dotsb a_N} \Pi^{\vec{\jmath}}}}$.
For the GQD of any N-partite state it is true that~\cite{Brag12}
    \begin{equation} \label{eq:gqd}
    \delta_g(a_1|\dotsb|a_N) \geq \sum_{k=1}^{N-1}
        { \delta_g(a_1\dotsb a_k|a_{k+1}) } \,.
    \end{equation}
For example, for the classical state ${\sigma^{AB}}$, if we
consider the partition ${a|\bar{a}|B}$, one has
${\delta_g(a|\bar{a}|B)\geq\delta_g(a|\bar{a})+%
\delta_g(a\bar{a}|B)}$, but
${\delta_g(a\bar{a}|B)=\delta_g(A|B)=0}$ and then
${\delta_g(a|\bar{a}|B) \geq \delta_g(a|\bar{a})}$. Alternatively,
we can consider the partition ${a|b|aux}$. We have then
${\delta_g(a|b|aux) \geq \delta_g(a|b) + \delta_g(ab|aux)}$.
Eq.~\eqref{eq:gqd} suggests that so as to obtain a monogamy
relation valid for dicord-like measures we need to appeal to
generalized multipartite measures that account for the partition's
internal structure.

\vspace{.6cm}\noindent

\noindent
Our examples strongly validate our initial thesis: classically
correlated states in Hilbert spaces of large-enough dimension
constitute quantum resources for undertaking processing
information tasks. We showed how Remote State Preparation and
Entanglement Distribution can be carried out using classical
states. Additionally, we exhibited two important aspects of
classical extensions of separable quantum states. First, from a
thermodynamic viewpoint concerned with extracting work from the
extension, we showed that the minimal extension of a given state
is related to the minimum possible work extraction from the
extended state. Second, we suggested that the possibility of
classically extending any separable state is strongly linked to
the non monogamous nature of the discord-type correlations. We
can only recover generalized monogamy relations by considering
genuine multipartite correlations.

\bibliographystyle{unsrt}

\end{document}